\documentclass [a4paper,12pt]{article}
\usepackage{amssymb}
\usepackage{graphicx}
\usepackage{amsmath}
\usepackage{fancybox}
\usepackage{color}
\usepackage{geometry}

\newcommand{\be}{\begin{equation}}
\newcommand{\ee}{\end{equation}}
\newcommand{\bea}{\begin{eqnarray}}
\newcommand{\eea}{\end{eqnarray}} 
\begin{document}
\begin{titlepage}

\vspace{1.0cm}
\begin{center}
{\LARGE\bf The implication of gauge-Higgs unification to the hierarchical fermion masses} 
\end{center}
\vspace{25mm}

\begin{center}
{\large
C. S. Lim$^*$
}
\end{center}
\vspace{1cm}

\centerline{{\it
$^*$
Department of Mathematics, Tokyo Woman's Christian University, Tokyo 167-8585, Japan }}
%
%
\vspace{2cm}
\centerline{\large\bf Abstract}
\vspace{0.5cm}

The observed hierarchical charged fermion masses of three generations seem to imply that these masses are originally universal for three generations and then get exponential suppression with ``quantized exponents" by some mechanism. We argue that such remarkable feature of hierarchical fermion mass spectrum may be naturally understood in the scenario of gauge-Higgs unification, where the universality of fermion masses is guaranteed by the fact that Higgs boson is originally gauge boson in this scenario and also the quantized exponents 
may be attributed to the well-known quantization condition of magnetic charge of the magnetic monopole placed inside the torus, as the extra dimensional space. Because of the presence of the magnetic monopole, we get a chiral theory and multiple Kaluza-Klein zero modes even if we introduce only a single 6-dimensional Weyl fermion. We present two types of three generation model, which succeed to realize the remarkable hierarchical mass spectrum.

\end{titlepage} 

\section{Introduction} 

There exist a few scenarios of physics beyond the standard model (BSM), which have been proposed as possible solutions to the well-known gauge hierarchy problem relying on some symmetries. The representative scenario is supersymmetry, whose concrete realization is minimal supersymmetric standard model (MSSM). In this paper we focus on the scenario of gauge-Higgs unification (GHU), where the origin of the Higgs boson is gauge boson. To be more precise, Higgs field is identified with the (Kaluza-Klein (KK) zero mode of ) extra space component of higher dimensional gauge field \cite{Manton:1979kb}, \cite{Hosotani}. A nice feature of this scenario is that, by virtue of higher dimensional local gauge symmetry, the quantum correction to the Higgs mass is UV-finite, thus opening a new avenue for the solution of the hierarchy problem and therefore the scenario 
of BSM \cite{Hatanaka:1998yp}. 

The MSSM and GHU also share a nice feature that the Higgs mass is ``calculable" as the result of the symmetry responsible for the protection of the Higgs mass, and the predicted Higgs masses are of the order of  $M_{W}$, to be consistent with the observed value $M_{H} = 125$ (GeV) by ATLAS \cite{Aad:2012tfa} and CMS \cite{Chatrchyan:2012xdj} experiments. This is basically because the self-couplings of Higgs field in these models are governed by gauge principle. For instance, the Higgs mass prediction including quantum correction made in a 6-dimensional (6D) GHU model is given in \cite{LMM2015}.           

Basic problem in the standard model is there is no principle to control the Higgs interactions, and therefore the strengths of Higgs interactions, such as Yukawa couplings and the self-coupling, can be arbitrary. Another motivation to study the GHU scenario is an expectation that it may naturally provide a mechanism to restrict the Higgs interactions, relying on the gauge principle.  

At the first glance, however, the GHU seems to be not a suitable scenario to discuss flavor physics; the Yukawa coupling is gauge coupling and therefore universal for all generations, at least to start with, since the origin of the Higgs boson is gauge boson. In this paper, however, we would like to point out that, on a contrary to this naive guess, the GHU scenario actually provides us with a natural framework to explain the remarkable hierarchical structure of the observed fermion masses \cite{Lim}.  

Before going into the detail of the statement, a comment on a mechanism of exponential suppression of the Yukawa coupling in 5D gauge theory is in order. In 5D gauge theories with an orbifold $S^{1}/Z_{2}$ as its compactified extra dimension, the mass term of ordinary type, $M_{b} \bar{\psi}\psi$ ($M_{b}$: bulk  mass) is not allowed even if it is Lorentz and gauge invariant. This is because, the $Z_{2}$ transformation for spinor fields is a sort of chiral transformation, $Z_{2}: \ \psi \ \to \ i\gamma_{5} \psi$ (for the case of simplified U(1) gauge theory). Nevertheless, so called ``$Z_{2}$-odd 
bulk mass term'' is still possible, being consistent with the $Z_{2}$ symmetry: 
\be 
\label{1.1} 
\epsilon (y) M_{b} \bar{\psi}\psi 
\ee 
where $y$ is the extra space coordinate and $\epsilon (y)$ is the ``sign function": $\epsilon (y) = 1 \ {\rm and} -1$ for positive and negative $y$, respectively. The remarkable consequence is that the mass term causes the localization of the mode function of KK zero mode of each Weyl fermion at different fixed points of the orbifold, $y = 0 \ {\rm and} \ \pm 
\pi R$ ($R$: the radius of the circle), depending on its chirality. As the Yukawa coupling is obtained by the overlap integral of the mode functions of left- and right-handed fermions, such localization results in an exponential suppression factor $\sim (\pi RM_{b}) e^{-\pi RM_{b}}$ of the Yukawa coupling, which is originally gauge coupling constant. It, however, also should be noticed that the bulk mass $M_{b}$ is just put by hand and its origin is unknown: there is no principle to fix the magnitude of $M_{b}$.

\section{What the GHU scenario implies to the hierarchical fermion masses}

As has been mentioned above, there is a remarkable hierarchical structure of the observed masses of charged fermions. Namely, if we plot $\log m_{f}$ ($m_{f}$: generic fermion mass) as the function of ``generation number" $N_{g} \ (N_{g} = 1, 2, 3)$, they seem to align along a straight line, roughly speaking, implying that $\log m_{f}$ is a linear function of $N_{g}$: 
\be 
\label{2.1} 
\log m_{f} = \gamma N_{g} + \delta \ \ \to \ \ m_{f} = e^{3\gamma + \delta}e^{- (3-N_{g}) \gamma} \ \ (\gamma, \ \delta: \ {\rm constants}). 
\ee 
This fact seems to suggest that the Yukawa couplings are originally all the same and universal for three generations and then get exponential suppression in some regular manner with ``quantized'' exponents. Let us note that $3 - N_{g} = 0, 1, 2$: integers.

What we would like to stress in this paper is that actually the GHU scenario just provides a natural framework to explain this interesting remarkable hierarchical structure:   

\noindent (A) As was already mentioned, in the GHU scenario Yukawa couplings are originally gauge coupling constant and therefore universal. This provides a good reasoning of the existence of the universal factor $e^{3\gamma + \delta}$ in (\ref{2.1}) in front of the exponential suppression factor. 

\noindent (B) The exponential suppression factor $e^{- (3-N_{g}) \gamma}$ may be also naturally realized by the factor  $e^{-\pi RM_{b}}$ due to the $Z_{2}$-odd bulk mass $M_{b}$. 

Thus only remaining task for us is to find a natural mechanism to realize the ``quantized" bulk mass $M_{b}$. One of the main purposes of this paper is to point out that such quantized bulk mass in 5D theory is naturally realized as the consequence of the presence of magnetic monopole placed inside a torus $T^{2}$ as the extra dimension of 6D GHU model.  The presence of the magnetic monopole yields background configuration of $A_{6}$, the 6th component of 6D gauge field, which is known to behave as a $Z_{2}$-odd bulk mass term from the viewpoint of 5D space-time. Importantly, the quantizaion of the ``bulk mass" is realized by the well-known Dirac's quantization condition of the magnetic charge of the monopole. 

Thus, the $Z_{2}$-odd bulk mass $M_{b}$ in 5D theory, whose origin was unknown, now acquires a physical interpretation as the background configuration of gauge field $A_6$ originating from the magnetic monopole, a topologically stable object. Also the magnitude of $M_{b}$ cannot be arbitrary and now is restricted by the quantization condition. Such replacement will be natural from the following physical reasons. 

\noindent (a) $Z_{2}$-odd bulk mass is ``parity-odd'' quantity in the sense that it changes sign under a sort of parity transformation in the extra dimension, $y \ \to \ -y$. So, it will be natural to replace it by the effect of magnetic monopole, which is also parity-odd quantity. 

\noindent (b) In (5D) GHU, Higgs field can be regarded as a sort of Aharonov-Bohm phase or the phase of Wilson-loop. So it may be natural to consider the effect of another possible magnetic physical object, magnetic monopole.   

\noindent (c) As fermionic matter field, we introduce 6D Weyl fermions with definite eigenvalue of 6D chiral operator $\Gamma_{7}$. Usually we still have non-chiral 4D theory. However, in our model, even if the extra space is just a torus $T^{2} = S^{1} \times S^{2}$ without ``orbifolding", it turns out that we actually get a chiral theory. Namely, concerning KK zero mode, only the mode function of either left- or right-handed Weyl fermion turns out to be normalizable. This may also be understood as a consequence of the parity violation mentioned above by the presence of the magnetic monopole. Similar mechanism has been pointed out in the scenario of ``magnetized extra dimension'' \cite{Cremades}. 

\noindent (d) As a bonus, it will be demonstrated below that even if we introduce only a single 6D Weyl fermion, we eventually get $M$ KK zero modes, where $M$ is the integer (assumed to be positive) appearing in the quantization condition of the magnetic monopole (see (\ref{3.5}) below). Such $M$ KK zero modes can be regarded as the existence of $M$ generations of each type of charged fermion. Again a similar result has been obtained in \cite{Cremades}.   

In the literature there are related works which discuss fermion mass spectrum by different approaches, e.g. by use of the magnetized extra dimension \cite{Kobayashi-1}, \cite{Kobayashi-2}, \cite{Kobayashi-3}, \cite{Kobayashi-4}, by use of point interactions in the extra dimension \cite{Fujimoto} or from the viewpoint of higher dimensional grand unified theories \cite{Kitano}, \cite{Haba}.

\section{The mechanism to realize the hierarchical fermion mass spectrum} 

\subsection{Minimal framework - U(1) gauge theory -}

In order to study the essence of the mechanism to realize the hierarchical fermion mass spectrum, we start with the discussion of minimal framework for the mechanism, i.e. 6D U(1) gauge theory, whose compactified extra dimension is  a torus $T^{2} = S^{1} \times S^{2}$. The radius of each $S^{1}$ of 5th and 6th dimension is denoted by $R_{5}, \ R_{6}$, respectively. The gauge coupling constant is written as $e$. We assume that a magnetic monopole with magnetic U(1) charge $g$ has been placed inside the torus, at the point of $y_{5} = \pm \pi R_{1}$, as is shown in Fig.1. (We denote extra-space coordinates by $y_{5}, \ y_{6}$.) To be more precise, the torus $T^{2}$ is assumed to have been embedded into higher dimensional (say three dimensional) extra space and the monopole is placed inside the torus, not on the surface of the torus. (Namely, monopole does not exist in our 6D world). The magnetic flux stemming from the magnetic monopole is assumed to penetrate through inside of the torus in two opposite directions and finally diverges towards the outside of the torus through its surface at the position of $y_{5} = 0$, as is seen in Fig.1.   


\begin{figure}[htbp]
 \begin{center}
  \includegraphics[width=120mm]{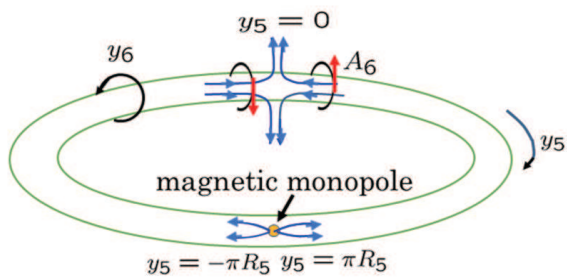}
 \end{center}
 \caption{The magnetic monopole placed inside of the torus}
\end{figure}

The U(1) gauge field is denoted as $A_{M} = (A_{\mu}, A_{5}, A_{6}) \ (\mu = 0, 1, 2, 3)$, and is treated as a classical background configuration to describe the magnetic flux stemming from the magnetic monopole. As matter field, we introduce a massless 6D Weyl fermion $\psi$ with U(1) charge $eQ$. 6D gamma matrices we adopt are written in the form of a direct product of the gamma matrices in 4D space-time and ``internal space" as follows: 
\be 
\label{3.1}
\Gamma^{\mu} = \gamma^{\mu} \otimes \sigma_{1}, \ \Gamma^{5} = i \gamma_{5} \otimes \sigma_{1}, \ 
\Gamma^{6} = -i I_{4} \otimes \sigma_{2}
\ee            
 
In this basis, 6D chiral operator is given as $\Gamma_{7} = I_{4} \otimes \sigma_{3}\ (I_{4}: 4\times 4 \ {\rm unit \ matrix})$ and therefore the 6D Weyl fermion $\psi$ with $+1$ eigenvalue of $\Gamma_{7}$ is just upper half of the full 8-component 6D spinor $\Psi$. Then, the 6D lagrangian for $\Psi$ just reduces to that for $\psi$: 
\be 
\label{3.2} 
i \bar{\Psi}D_{M}\Gamma^{M} \Psi = \bar{\psi} \{ i ( D_{\mu} \gamma^{\mu} + D_{5}i \gamma_{5}) + i D_{6} \}   \psi.   
\ee 
We thus realize that $i D_{6}$, especially $eQ A_{6}$ behaves as a mass term for $\psi$ from 5D point of view. 

Since our mechanism to realize the desirable hierarchical mass spectrum is based on the considerations in 5D space-time, 
we assume 
\be 
\label{3.3} 
R_{5} \gg R_{6} 
\ee 
to recover 5D point of view. Thus, the torus becomes a thin tube. Then, as is shown in Fig.1 the magnetic flux stemming from the monopole, assumed to be confined inside of the torus, has a direction along the cycle of the 5th dimension. The direction of the magnetic flux changes at the origin of the coordinate $y_5$. Accordingly, $A_{6}$ also changes its sign there, behaving as 
\be 
\label{3.4} 
A_{6} = \epsilon (y_5) \frac{g}{4\pi R_{6}}, 
\ee 
so that, by use of the Stokes' theorem, $\oint A_{6} dy_{6} = \epsilon (y_5) \frac{g}{2}$. Clearly, this configuration of $A_{6}$ just mimics the $Z_{2}$-odd bulk mass in (\ref{1.1}). In addition, the magnetic charge $g$ cannot be arbitrary: it should satisfy the well-known Dirac's quantization condition 
\be 
\label{3.5} 
eQg = 2\pi M \ \ (M: {\rm integer}), 
\ee 
which hence implies  
\be 
\label{3.6} 
eQA_{6} = \epsilon (y_5) \frac{M}{2R_{6}} \ \ (M: {\rm integer}).
\ee 
In this way, the necessary quantized bulk mass is (effectively) realized by the introduction of the magnetic monopole. 

Here is a comment on how the presence of the magnetic monopole leads to a chiral 4D theory. In a suitable choice of the 6D gamma matrices (different from those given in (\ref{3.1})), $\Gamma_{7} = \gamma_{5} \otimes \sigma_{3}$, the product of 4D chiral operator and the chiral operator of 2D extra dimension. Since the ``parity transformation" in the 2D extra space, $y_{5} \ \to \ -y_{5}$, changes the chirality of the extra dimension (the eigenvalue of $\sigma_{3}$), the presence of the magnetic monopole with odd parity, which is clearly seen in (\ref{3.6}), causes asymmetry between two chiralities in the extra dimension. As the matter field $\psi$ is 6D Wely fermion with definite eigenvalue of $\Gamma_{7}$, this means that the presence of the magnetic monopole also causes the asymmetry between two 4D chiralities, thus leading to a chiral 4D theory. We will see this really is the case in the discussion of the mode function of KK zero mode below.

\subsection{Mode functions for KK zero modes} 

In order to see whether the Yukawa coupling really gets the desirable exponential suppression as we expected, it is necessary to have a concrete form of the mode functions for the KK zero modes of fermion $\psi$. Writing a KK zero mode of 4D Weyl fermion, say left-handed fermion, as $\psi^{(0)}_{L} (x^{\mu}, y_{5}, y_{6}) = \psi^{(0)}_{L}(x^{\mu}) f_{L}(y_{5}, y_{6})$, the mode function 
$f_{L}(y_{5}, y_{6})$ should satisfy 
\be 
\label{4.1} 
(-D_{5} \gamma_{5} + i D_{6}) \psi^{(0)}_{L} (x^{\mu}, y_{5}, y_{6}) = 0 \ \to \ 
(D_{5} + i D_{6}) f_{L}(y_{5}, y_{6}) = 0.  
\ee 
The classical (background) configuration of the extra space components of gauge field due to the magnetic flux stemming from the magnetic monopole is 
\be 
\label{4.2} 
A_{5} = 0, \ \ A_{6} = \epsilon (y_5) \frac{g}{4\pi R_{6}}. 
\ee 
Then the differential equation for the mode function $f_{L}(y_{5}, y_{6})$ reads as 
\be 
\label{4.3} 
\{ \partial_{5} + i (\partial_{6} - ieQ \epsilon (y_5) \frac{g}{4\pi R_{6}}) \} f_{L}(y_{5}, y_{6}) = 0.
\ee 

Since there is no magnetic monopole put at $y_{5} = 0$, the mode function is supposed to be continuous there: the mode function can be written as 
\be 
\label{4.4} 
f_{L}(y_{5}, y_{6}) = e^{- \frac{eQg}{4\pi R_{6}}|y_{5}|} \hat{f}_{L}(y_{5}, y_{6})
\ee 
where $\hat{f}_{L}(y_{5}, y_{6})$ is a continuous function for $- \pi R_{5} < y_{5} < \pi R_{5}$, satisfying 
\be 
\label{4.5}
(\partial_{5} + i \partial_{6}) \hat{f}_{L}(y_{5}, y_{6}) = 0. 
\ee 
Thus $\hat{f}_{L}(y_{5}, y_{6})$ can be arbitrary function of $y_{5} + iy_{6}$, and assuming the periodic boundary condition along the cycle of 6th dimension, $f_{L}(y_{5}, y_{6}+2\pi R_{6}) = f_{L}(y_{5}, y_{6})$ the mode function can be written in a form of Fourier expansion:  
\be 
\label{4.6} 
f_{L}(y_{5}, y_{6}) = e^{- \frac{eQg}{4\pi R_{6}}|y_{5}|} \sum_{n} c_{n} e^{i\frac{n}{R_{6}}(y_{6}-iy_{5})},  
\ee
with coefficients $c_{n}$. (We will think about the possibility of ``twisted boundary condition" later.) 

Even though the mode function is continuous for $- \pi R_{5} < y_{5} < \pi R_{5}$, there appears a discontinuity of the mode function at $y_{5} = \pm \pi R_{5}$ because of the presence of the magnetic monopole. The gap $\frac{g}{2\pi R_{6}}$ of $A_{6}$ in the both sides of $y_{5} = \pm \pi R_{5}$, may be resolved by a gauge transformation in the region of, e.g., $y_{5} < 0$ with a transformation parameter 
$\frac{eQg}{2\pi R_{6}}y_{6}$. Since the mode functions in both sides should be connected by this gauge transformation, they should satisfy a relation 
\be 
\label{4.7}
f_{L}(\pi R_{5}, y_{6}) = e^{i \frac{eQg}{2\pi R_{6}}y_{6}}f_{L}(- \pi R_{5}, y_{6}).  
\ee 
For the extra phase factor not to spoil the periodicity of the mode function along 6th dimension, the multiple of ``electric" charge $eQ$ and the magnetic charge $g$ should satisfy a condition $eQg = 2\pi M \ (M: \ {\rm integer})$, which is nothing but the 
well-known quantization condition (\ref{3.5}). In terms of the integer $M$, the mode function is simplified as 
\be 
\label{4.8} 
f_{L}(y_{5}, y_{6}) = e^{- \frac{M}{2 R_{6}}|y_{5}|} \sum_{n} c_{n} e^{i\frac{n}{R_{6}}(y_{6}-iy_{5})},
\ee 
which is subject to the condition (\ref{4.7}) in terms of $M$, 
\be 
\label{4.9}
f_{L}(\pi R_{5}, y_{6}) = e^{i \frac{M}{R_{6}}y_{6}}f_{L}(- \pi R_{5}, y_{6}). 
\ee 
Substituting (\ref{4.8}) in the condition (\ref{4.9}), we get a condition to be satisfied by the coefficients $c_{n}$, 
\be 
\label{4.10} 
\sum_{n} c_{n} e^{\frac{n \pi R_{5}}{R_{6}}} e^{i\frac{n}{R_{6}}y_{6}} = \sum_{n} c_{n} e^{- \frac{n \pi R_{5}}{R_{6}}} e^{i\frac{n + M}{R_{6}}y_{6}}.  
\ee 
Comparing the coefficients of the same Fourier mode in both sides of (\ref{4.10}), we get a recursion formula, 
\be 
\label{4.11} 
c_{n + M} = e^{- \frac{(2n + M) \pi R_{5}}{R_{6}}} c_{n}.  
\ee 
This relation leads to an important conclusion that for a fixed $j \ (j = 0,1, \cdots, |M|-1)$, coefficients $c_{j + rM} \ (r: {\rm integer})$ all depend on $c_{j}$, 
\be 
\label{4.12} 
c_{j + rM} = e^{-\frac{\pi R_{5}}{R_{6}}(Mr^{2} + 2jr)} c_{j},  
\ee 
which means there are $|M|$ independent KK zero modes characterized by $j$ (if they are ever normalizable).  This may be regarded as the presence of $|M|$ generations of fermion of some specific type of fermion (up-type quark, etc.). From (\ref{4.12}), for a fixed $M$ the mode function of $j$-th KK zero mode $(j = 0,1, \cdots, |M|-1)$ is obtained as 
\bea 
f_{L}^{(M,j)}(\hat{y_{5}}, \hat{y_{6}}) &=& c^{(M,j)} \sum_{r=-\infty}^{\infty} e^{-\frac{\pi R_{5}}{R_{6}}M(r + \frac{j}{M})^{2}} 
e^{\frac{\pi R_{5}}{R_{6}}M\{ (r + \frac{j}{M})\hat{y_{5}} - \frac{1}{2}|\hat{y}_{5}| \}} e^{i\pi M(r + \frac{j}{M})\hat{y_{6}}}  \nonumber \\ 
&=& c^{(M,j)} e^{-\frac{\pi R_{5}}{R_{6}}M \{ (\frac{j}{M})^{2} - \frac{j}{M}(\hat{y}_{5} + i \frac{R_{6}}{R_{5}}\hat{y}_{6}) + \frac{|\hat{y}_{5}|}{2} \}} \nonumber \\ 
&& \cdot \theta_{3} (i\frac{R_{5}}{R_{6}}M (\frac{j}{M} - \frac{\hat{y}_{5}+i \frac{R_{6}}{R_{5}}\hat{y}_{6}}{2}) \ | \ i\frac{R_{5}}{R_{6}}M),    
\label{4.13} 
\eea 
where dimensionless coordinates $\hat{y_{5}} = \frac{y_{5}}{\pi R_{5}}, \ \hat{y_{6}} = \frac{y_{6}}{\pi R_{6}}$ have been introduced with 
$|\hat{y}_{5,6}| \leq 1$. The overall factor $c^{(M,j)}$ should be fixed by the normalization condition of the mode function. In the second line of (\ref{4.13}), $\theta_{3}$ is one of the Jacobi theta functions, defined by  
\be 
\label{4.14} 
\theta_{3} (\nu | \tau) \equiv \sum_{r = -\infty}^{\infty} q^{r^{2}} e^{i\pi 2r\nu} \ \ (q = e^{i\pi \tau}). 
\ee

As the matter of fact, the infinite series in (\ref{4.13}) is convergent only for $M > 0$, which we assume to be the case henceforth. In this convergent infinite series, under the assumption (\ref{3.3}), it turns out that only a few terms provide potentially important contributions. To see this, let us consider the following infinite sum,   
\bea
\sum_{r = -\infty}^{\infty} e^{- \alpha (r+\beta)^{2}} &=& e^{- \alpha \beta^{2}} \theta_{3}(i\frac{\alpha \beta}{\pi} | i \frac{\alpha}{\pi}) \nonumber \\ 
&=& e^{- \alpha \beta^{2}} \Pi_{l = 1}^{\infty} (1 - e^{-2l \alpha}) \cdot \Pi_{n = 1}^{\infty} \{ 1 + 2 e^{-(2n-1)\alpha} \cosh (2\alpha \beta) + e^{-(4n-2)\alpha} \} \nonumber \\ 
&=& e^{- \alpha \beta^{2}} \Pi_{l = 1}^{\infty} (1 - e^{-2l \alpha}) \cdot \Pi_{n = 1}^{\infty} \{ (1 + e^{-(2n-1-2\beta)\alpha}) \cdot (1 + e^{-(2n-1+2\beta)\alpha}) \}   \nonumber \\  
&=& e^{- \alpha \beta^{2}} \{ \Pi_{l = 1}^{\infty} (1 - e^{-2l \alpha}) \} \cdot (1 + e^{(-1+2\beta) \alpha})(1 + e^{(-3+2\beta)\alpha}) \cdots \nonumber \\ 
&\cdot & (1 + e^{(-1-2\beta) \alpha})(1 + e^{(-3-2\beta)\alpha}) \cdots . 
\label{4.15} 
\eea
The case of our interest corresponds to $\alpha = \frac{\pi R_{5}}{R_{6}}M, \ \beta = \frac{j}{M} - \frac{\hat{y}_{5}+i \frac{R_{6}}{R_{5}}\hat{y}_{6}}{2}$. Let us note $-\frac{1}{2} \leq {\rm Re} \ \beta = \frac{j}{M} - \frac{\hat{y}_{5}}{2} < \frac{3}{2}$. Under this condition, when $\alpha \gg 1$ almost all factors in the infinite products in (\ref{4.15}) is well-approximated by 1, except for $(1 + e^{(-1+2\beta) \alpha})$. 
Thus we conclude that  
\be 
\label{4.16} 
\sum_{r = -\infty}^{\infty} e^{- \alpha (r+\beta)^{2}} \simeq  e^{- \alpha \beta^{2}}(1 + e^{(-1+2\beta) \alpha}) = e^{- \alpha \beta^{2}} + e^{- \alpha (-1+\beta)^{2}}. 
\ee 
We realize that this just means that in the sum (\ref{4.15}), only the terms of $r = 0, \ -1$ give potentially non-negligible contributions for $-\frac{1}{2} \leq {\rm Re} \ \beta < \frac{3}{2}$, as we naively expect.  

We thus obtain well-approximated expression for the mode function of KK zero mode: 
\bea 
&& f^{(M,j)}_{L}(\hat{y_{5}}, \hat{y_{6}}) \nonumber \\ 
&\simeq& c^{(M,j)} \{ e^{-\frac{\pi R_{5}}{R_{6}}\frac{j^{2}}{M}} 
e^{- \frac{\pi R_{5}}{R_{6}}\{ -j\hat{y_{5}} + \frac{M}{2}|\hat{y}_{5}| \}} e^{i\pi j \hat{y}_{6}}   
+ e^{-\frac{\pi R_{5}}{R_{6}}\frac{(M-j)^{2}}{M}} 
e^{-\frac{\pi R_{5}}{R_{6}}\{ (M - j)\hat{y_{5}} + \frac{M}{2}|\hat{y}_{5}| \}} e^{-i\pi (M-j) \hat{y}_{6}} \} 
 \nonumber \\ 
&\simeq&  
\begin{cases} 
\frac{1}{\sqrt{\pi} R_{6}} \sqrt{\frac{\frac{M^{2}}{4} - j^{2}}{M}} \cdot 
\left( \theta (\hat{y}_{5})e^{- \frac{\pi R_{5}}{R_{6}}(\frac{M}{2}-j) \hat{y_{5}}} + \theta (-\hat{y}_{5})e^{ \frac{\pi R_{5}}{R_{6}}(\frac{M}{2}+ j) \hat{y_{5}}} \right) \cdot e^{i\pi j \hat{y}_{6}} \\ 
 \ \  \text{(for \ $0 \leq j < \frac{M}{2}$)} \\ 
\frac{1}{\sqrt{\pi} R_{6}} \sqrt{\frac{\frac{M^{2}}{4} - (M-j)^{2}}{M}} \cdot 
\left( \theta (\hat{y}_{5})e^{- \frac{\pi R_{5}}{R_{6}}(\frac{3M}{2}-j) \hat{y_{5}}} + \theta (-\hat{y}_{5})e^{ \frac{\pi R_{5}}{R_{6}}(j - \frac{M}{2}) \hat{y_{5}}}\right) \cdot e^{-i\pi (M-j) \hat{y}_{6}} \\ 
 \ \  \text{(for \ $\frac{M}{2} < j < M$)} \\  
\frac{1}{2\pi \sqrt{R_{5}R_{6}}} \cdot 
\left( \theta (\hat{y}_{5}) e^{i\pi \frac{M}{2} \hat{y}_{6}} + \theta (-\hat{y}_{5}) e^{-i\pi \frac{M}{2} \hat{y}_{6}} \right) \\ 
 \ \ \text{(for \ $j = \frac{M}{2}$, $M$: even)},  \\  
\end{cases} 
\label{4.17}  
\eea 
where we have fixed the normalization factor $c^{(M, j)}$, so that 
\be 
\label{4.18} 
\pi^{2} R_{5}R_{6} \int_{-1}^{1} d\hat{y}_{5} \int_{-1}^{1} d\hat{y}_{6} |f^{(M,j)}_{L}(\hat{y_{5}}, \hat{y_{6}})|^{2} = 1. 
\ee 

(\ref{4.17}) clearly shows that the mode functions behave as exponential functions with ``quantized" exponents, with $M, \ j$ being integers, and the left-handed KK zero modes are localized at $y_{5} = 0$. This mimics the localization due to the $Z_{2}$-odd bulk mass $M_{b}$ in 5D theory, and suggests that the Yukawa couplings are also suppressed by the exponential factor with quantized exponents. 

We now turn to the KK zero modes of 4D right-handed fermion \\ 
$\psi^{(0)}_{R} (x^{\mu}, y_{5}, y_{6}) = \psi^{(0)}_{R}(x^{\mu}) f_{R}(y_{5}, y_{6})$, which satisfies 
\be 
\label{4.19} 
(-D_{5} \gamma_{5} + i D_{6}) \psi^{(0)}_{R} (x^{\mu}, y_{5}, y_{6}) = 0 \ \to \ 
\{ - \partial_{5} + i (\partial_{6} - ieQ \epsilon (y_5) \frac{g}{4\pi R_{6}}) \} f_{R}(y_{5}, y_{6}) = 0.   
\ee 
The solution $f_{R}^{(M,j)}(\hat{y_{5}}, \hat{y_{6}})$ of this equation is easily understood to be given by 
\be 
\label{4.20} 
f_{R}^{(M,j)}(\hat{y_{5}}, \hat{y_{6}}) = f_{L}^{(-M,j)}(-\hat{y_{5}}, \hat{y_{6}}). 
\ee 
As is seen from (\ref{4.13}), however, since $M > 0$ the infinite series in (\ref{4.20}) is divergent and hence the mode function is not normalizable. Thus, we conclude for a 6D Weyl (or equivalently 4D Dirac) fermion $\psi$ with the positive integer $M$, only left-handed fermion has $M$ KK zero modes. As the result, 
we have succeeded in realizing a chiral theory, even though we have not adopted any orbifolding. For $\psi$ with integer $-M$, the situation is just opposite and only right-handed fermion has $M$ KK zero modes, whose mode functions are given by 
\be 
\label{4.21} 
f_{R}^{(-M,j)}(\hat{y_{5}}, \hat{y_{6}}) = f_{L}^{(M,j)}(-\hat{y_{5}}, \hat{y_{6}}). 
\ee 
which are clearly normalizable.

\subsection{Overlap integrals of mode functions} 

To get Dirac masses for fermions through spontaneous gauge symmetry breaking, Yukawa couplings are needed, which are given by the product of the gauge coupling constant and the overlap integrals of fermion's mode functions for the KK zero modes of both chiralities (assuming that the mode function of the KK zero mode of Higgs field, to be identified with that of $A_{5}$ or $A_{6}$ in the GHU scenario, is just a constant).

We now encounter a problem. Namely, as is seen from (\ref{4.17}) and (\ref{4.21}), both mode functions of left- and right-handed fermion are localized at the same point $y_{5} = 0$, in clear contrast to the case of orbifold compactification with $Z_{2}$-odd bulk mass, where each KK zero mode is localized at different fixed point depending on its chirality, leading to the exponential suppression of the Yukawa coupling. Thus, the overlap integral does not acquire the exponential suppression in the present form of mode function. In fact, for $0 \leq j < \frac{M}{2}$, for instance, 
\bea 
\label{5.1} 
&& \pi^{2} R_{5}R_{6} \int_{-1}^{1} d\hat{y}_{5} \int_{-1}^{1} d\hat{y}_{6} f_{L}^{(M,j)}(\hat{y_{5}}, \hat{y_{6}})^{\ast} f_{R}^{(-M,j)}(\hat{y_{5}}, \hat{y_{6}}) \nonumber \\ 
&& = 2 \pi^{2} R_{5}R_{6} \int_{0}^{1} d\hat{y}_{5} \int_{-1}^{1} d\hat{y}_{6} f_{L}^{(M,j)}(\hat{y_{5}}, \hat{y_{6}})^{\ast} f_{L}^{(M,j)}(-\hat{y_{5}}, \hat{y_{6}})  \nonumber \\ 
&& \simeq \frac{M^{2} - 4j^{2}}{M^{2}}, 
\eea 
which is just power suppressed. This problem is not resolved even if we adopt the 6D Weyl fermion with negative eigenvalue of $\Gamma_{7}$, since this change causes $D_{6} \to - D_{6}$, which is equivalent to $\partial_{5} \to - \partial_{5}$ and therefore the change of 4D chirality.

\subsection{U(1)$\times$U(1) gauge theory and exponentially suppressed Yukawa coupling}   

Possible solution to the problem mentioned in the previous subsection is to prepare another monopole at the opposite side of the torus, i.e. at $y_{5} = 0$ whose magnetic charge is associated with anther U(1) gauge symmetry, independent of the original U(1) symmetry, and fermion to yield right-handed fermion feels this newly introduced magnetic monopole, so that the right-handed fermion localizes at $y_{5} = \pm \pi R_{5}$. 

In order to make this idea concrete, we consider a gauge theory with a little extended gauge symmetry: 6D U(1)$_{1} \times$ U(1)$_{2}$ gauge theory, whose gauge coupling constants are assumed to be the same for brevity, i.e. $e$. We place two independent magnetic monopoles at $y_{5} = \pm \pi R_{5}$ and $y_{5} = 0$, carrying magnetic charges of U(1)$_{1}$ and U(1)$_{2}$, respectively. The magnetic charges $g_{1}, \ g_{2}$ of two magnetic monopoles are assumed to be the same, again for brevity: 
\be 
\label{5.2}  
g_{1} = g_{2} = \frac{2\pi}{e}. 
\ee 

As the matter fields, we introduce a pair of 6D Weyl fermions $\psi_{1}, \ \psi_{2}$, both having the same eigenvalue $+1$ of $\Gamma_{7}$. Their ``charge" assignments are as follows: 
\be 
\label{5.3}
\psi_{1} : \ (M, 0), \ \ \psi_{2}: \ (0, -M),  
\ee 
where $(M, 0), \ (0, -M)$ ($M \geq 0$: integer) denote the charges of (U(1)$_1$, \ U(1)$_2$). 
Let us note that under (\ref{5.2}), the quantization condition (\ref{3.5}) means $Q = M$. 
Then $\psi_{1}$ has $M$ independent KK zero modes of 4D left-handed fermion, while $\psi_{2}$ has $M$ independent KK zero modes of 4D right-handed fermion, localized at $y_{5} = 0$ and $y_{5} = \pm \pi R_{5}$, respectively. Thus we now acquire the desirable exponentially suppressed Yukawa couplings, and therefore fermion masses, as we will see below.  

In this model, the mode functions of KK zero modes for the right-handed fermion with charge $-M$ is obtained by 
 $\hat{y}_{5} \to - \hat{y}_{5}$ followed by the translation in the 5th dimension, 
$\hat{y}_{5} \to \hat{y}_{5} - 1$ of the original model functions of the left-handed fermion given in (\ref{4.17}). To be concrete, the mode functions are given as  
\bea 
&& f_{R}^{(-M,j)} (\hat{y_{5}}, \hat{y_{6}}) \nonumber \\ &\simeq&  
\begin{cases} 
\frac{1}{\sqrt{\pi} R_{6}} \sqrt{\frac{\frac{M^{2}}{4} - j^{2}}{M}} \cdot 
\left( \theta (\hat{y}_{5})e^{- \frac{\pi R_{5}}{R_{6}}(\frac{M}{2}-j) (1 - \hat{y_{5}})} + \theta (-\hat{y}_{5})e^{ - \frac{\pi R_{5}}{R_{6}}(\frac{M}{2}+ j) (1 + \hat{y_{5}})} \right) \cdot e^{i\pi j \hat{y}_{6}} \\ 
 \ \  \text{(for \ $0 \leq j < \frac{M}{2}$)} \\ 
\frac{1}{\sqrt{\pi} R_{6}} \sqrt{\frac{\frac{M^{2}}{4} - (M-j)^{2}}{M}} \cdot 
\left( \theta (\hat{y}_{5})e^{- \frac{\pi R_{5}}{R_{6}}(\frac{3M}{2}-j) (1 - \hat{y_{5}}) } + \theta (-\hat{y}_{5})e^{- \frac{\pi R_{5}}{R_{6}}(j - \frac{M}{2}) (1 + \hat{y_{5}})} \right) \cdot e^{-i\pi (M-j) \hat{y}_{6}} \\ 
 \ \  \text{(for \ $\frac{M}{2} < j < M$)} \\  
\frac{1}{2\pi \sqrt{R_{5}R_{6}}} \cdot 
\left( \theta (\hat{y}_{5}) e^{i\pi \frac{M}{2} \hat{y}_{6}} + \theta (-\hat{y}_{5}) e^{-i\pi \frac{M}{2} \hat{y}_{6}} \right) \\ 
 \ \ \text{(for \ $j = \frac{M}{2}$, $M$: even)}.  \\  
\end{cases} 
\label{5.4}  
\eea  

Hence the overlap integral of the mode functions for left- and right-handed fermions is calculated to be 
\bea 
\label{5.5} 
&& \pi^{2} R_{5}R_{6} \int_{-1}^{1} d\hat{y}_{5} \int_{-1}^{1} d\hat{y}_{6} f_{L}^{(M,j)}(\hat{y_{5}}, \hat{y_{6}})^{\ast} f_{R}^{(-M,j)}(\hat{y_{5}}, \hat{y_{6}}) \nonumber \\ 
&\simeq &   
\begin{cases} 
\pi \frac{R_{5}}{R_{6}} M e^{- \frac{\pi R_{5}}{R_{6}}\frac{M}{2}} \ \  \text{(for \ $j = 0$)} \\ 
2\pi \frac{R_{5}}{R_{6}} \frac{\frac{M^{2}}{4}-j^{2}}{M} e^{- \frac{\pi R_{5}}{R_{6}}(\frac{M}{2}-j)} \ \  \text{(for \ $0 < j < \frac{M}{2}$)} \\ 
2\pi \frac{R_{5}}{R_{6}} \frac{\frac{M^{2}}{4} - (M-j)^{2}}{M} e^{- \frac{\pi R_{5}}{R_{6}}(j - \frac{M}{2})} \ \  \text{(for \ $\frac{M}{2} < j < M$)} \\  
1 \ \ \text{(for \ $j = \frac{M}{2}$, $M$: even)}.   \\  
\end{cases}   
\eea 
As we expected, we now have realized the desirable exponentially suppressed Yukawa couplings with quantized exponents, behaving as 
\be 
\label{5.6} 
\propto e^{- \frac{\pi R_{5}}{R_{6}}|\frac{M}{2}-j|}. 
\ee

\section{Three generation model} 

We are now ready to discuss the model of our real interest, i.e. three generation model. 
Obviously, the simplest possibility to get three generations is to introduce one pair of 6D Weyl fermions with charges $(3, 0), \ (0, -3)$ ($M = 3$ in (\ref{5.3})). Unfortunately, however, this model does not work. Namely, setting $M = 3$ in (\ref{5.5}) we find that the Yukawa couplings for three KK zero modes corresponding to $j = 0, \ 1, \ 2$ behave as $e^{- \frac{\pi R_{5}}{R_{6}}\frac{3}{2}}, \ e^{- \frac{\pi R_{5}}{R_{6}}\frac{1}{2}}, \ e^{- \frac{\pi R_{5}}{R_{6}}\frac{1}{2}}$, respectively, if we ignore the numerical factors in front of the exponential suppression factors. (This behavior also can be read off from (\ref{5.6})). It means that there appears a degeneracy of fermion masses, which apparently contradicts with the observed fermion mass spectrum of charged fermions. (Possible relevance of this model to the neutrino masses will be discussed in the summary discussion). We consider two possible toy models below, which can evade this difficulty.

\subsection{$2 + 1$ model} 

It is easily known from  (\ref{5.5}) or (\ref{5.6}) that the degeneracy mentioned above always happens for $M \geq 3$. So (almost) unique  possibility to get a satisfactory three generation model without degeneracy is to consider, say ``$2 + 1$"  model. Namely, we introduce one pair of fermions with $M = 2$ and another pair of fermions with $M = 1$. The pair with $M = 2$ is known from (\ref{5.5}) to provide exponential suppression factors  
\be 
\label{6.1} 
e^{- \frac{\pi R_{5}}{R_{6}}} \ (j = 0), \ \ 1 \ (j = 1),    
\ee
while the pair with $M = 1$ provides exponential suppression factor  
\be 
\label{6.2} 
e^{- \frac{\pi R_{5}}{R_{6}}\frac{1}{2}} \ (j = 0).     
\ee 
Thus three mass eigenvalues of fermion denoted by $m_{1}, \ m_{2}, \ m_{3} \ (m_{1} < m_{2} < m_{3})$ are identified as  
\be 
\label{6.2'} 
m_{1} \propto e^{- \frac{\pi R_{5}}{R_{6}}}, \ m_{2} \propto e^{- \frac{\pi R_{5}}{R_{6}}\frac{1}{2}}, \ m_{3} \propto 1.  
\ee 
Fortunately, the ratios of these mass eigenvalues (ignoring the numerical factors) satisfy a relation, which is exactly what we need to explain the observed remarkable hierarchical structure of charged fermion masses, 
\be 
\label{6.3} 
\frac{m_{3}}{m_{2}} = \frac{m_{2}}{m_{1}} = 
e^{\frac{1}{2} \frac{\pi R_{5}}{R_{6}}}.  
\ee 
Namely, this model predicts $\gamma = \frac{1}{2} \frac{\pi R_{5}}{R_{6}}$ in (\ref{2.1}). 

Since (\ref{6.3}) is based on the approximate form ignoring the numerical factors in (\ref{5.5}), actually when $\log m_{f}$ are plotted as a function of the generation number, there will 
be a slight deviation from a straight line implied by $\frac{m_{3}}{m_{2}} = \frac{m_{2}}{m_{1}}$. It will be interesting to see whether the slight deviation from the straight line also seen in the observed fermion mass spectrum can be attributed to such an effect. If we take into account the numerical factors appearing in (\ref{5.5}), 
the two ratios are actually given as $m_{2}/m_{1} = e^{\frac{\pi R_{5}}{R_{6}}\frac{1}{2}}/2, \ m_{3}/m_{2} = e^{\frac{\pi R_{5}}{R_{6}}\frac{1}{2}}/\{ \pi (R_{5}/R_{6})\}$. Thus the following ``double ratio" is no longer unity but is modified to  
\be 
\label{6.4} 
\frac{\left( \frac{m_{3}}{m_{2}} \right)}{\left( \frac{m_{2}}{m_{1}} \right)} 
= \frac{1}{\log \{2 (\frac{m_{2}}{m_{1}}) \}}. 
\ee 
If we take charged leptons as an example of charged fermion in order to fix the factor $\log \{2 (\frac{m_{2}}{m_{1}}) \}$ in the right hand side of (\ref{6.4}), identifying $m_{1} = m_{e} = 0.5$ MeV, $m_{2} = m_{\mu} = 106$ MeV, the double ratio is predicted to be 
\be 
\label{6.5} 
\frac{\left( \frac{m_{3}}{m_{2}} \right)}{\left( \frac{m_{2}}{m_{1}} \right)} 
= 0.17, 
\ee 
while the corresponding observed value of the double ratio for the charged lepton is  
\be 
\label{6.6}
\frac{\left( \frac{m_{\tau}}{m_{\mu}} \right)}{\left( \frac{m_{\mu}}{m_{e}} \right)} = 0.075,   
\ee 
which at least shows the same tendency as (\ref{6.5}), in the sense that $m_{3}/m_{2}$ is smaller than $m_{2}/m_{1}$. The difference of roughly factor 2 between (\ref{6.5}) and (\ref{6.6}) may not be remarkable in the plot of $\log m_{f}$. 

Though the detail is not shown here, we also have studied the case of up-type quarks, and have found a difference of roughly factor 2 between the predicted and observed double ratio, again. For down-type quarks, the agreement is not good.

\subsection{Model with twisted boundary condition}

Another possibility ot evade the problem of the degeneracy of mass eigenvalues is to consider ``twisted boundary condition". Let us note that the boundary condition along the circle of the 6th dimension of $\hat{f}_{L}(y_{5}, y_{6})$ satisfying (\ref{4.5}) needs not to be periodic boundary condition, but can be ``twisted". Namely, we can generalize the solution of (\ref{4.5}) to the following, with $0 \leq a < 1$ denoting the twisted boundary condition, 
\be 
\label{6.7} 
f_{L}(y_{5}, y_{6}) = e^{- \frac{eQg}{4\pi R_{6}}|y_{5}|} \sum_{n} c_{n} e^{i\frac{n + a}{R_{6}}(y_{6}-iy_{5})}.   
\ee 
The parameter $a$ can be regarded as a constant shift of the KK modes, so the $M$ KK zero modes are now given by (\ref{4.13}), where $j \ (j = 0,1, \cdots , M-1)$ is replaced by $j + a \ (0 \leq a < 1)$. So now $j + a$ can take arbitrary real number in the range of $0 \leq j + a < M$. Accordingly, the overlap integral (\ref{5.5}) is modified into 
\bea 
\label{6.8} 
&& \pi^{2} R_{5}R_{6} \int_{-1}^{1} d\hat{y}_{5} \int_{-1}^{1} d\hat{y}_{6} f_{L}^{(M,j+a)}(\hat{y_{5}}, \hat{y_{6}})^{\ast} f_{R}^{(-M,j+a)}(\hat{y_{5}}, \hat{y_{6}}) \nonumber \\ 
&\propto&   
\begin{cases} 
e^{- \frac{\pi R_{5}}{R_{6}}(\frac{M}{2}-(j + a))} \ \  \text{(for \ $0 \leq j + a \leq \frac{M}{2}$)} \\ 
e^{- \frac{\pi R_{5}}{R_{6}}(j + a - \frac{M}{2})} \ \  \text{(for \ $\frac{M}{2} < j + a < M$)} \\  
\end{cases}.    
\eea 

We now discuss whether the desirable hierarchical structure of the fermion masses can be obtained by the presence of $a$ for the three generation model, where only one pair of fermions with $M = 3$ is introduced. 
At first glance, it seems that we have to consider two cases, $0 \leq a < \frac{1}{2}$ and $\frac{1}{2} \leq a < 1$. It, however, is known that they are not independent as long as the overlap integral is concerned. This is because from (\ref{4.13}) a relation $f_{L}^{(M,j+a)}(-\hat{y_{5}}, -\hat{y_{6}}) = f_{L}^{(M, (M-j-1)+(1-a))}(\hat{y_{5}}, \hat{y_{6}})$ holds, while the overlap integral is invariant under the change of variables, $\hat{y_{5}} \to -\hat{y_{5}}, \ \hat{y_{6}} \to -\hat{y_{6}}$. Thus we need to consider only the  case of $0 \leq a \leq \frac{1}{2}$. In this case, the exponential suppression factor  given in (\ref{6.8}) for each of $j = 0, 1, 2$ are 
\be 
\label{6.9} 
e^{- \frac{3-2a}{6}\alpha} \ (j = 0), \ \ e^{- \frac{1-2a}{6}\alpha} \ (j = 1), \ \ e^{- \frac{1+2a}{6}\alpha} \ (j = 2),  
\ee
respectively, where $\alpha \equiv \frac{\pi R_{5}}{R_{6}}M = 3\frac{\pi R_{5}}{R_{6}}$. As $1-2a \leq 1+2a \leq 3-2a$  (for $0 \leq a \leq \frac{1}{2}$), the condition to realize the observed remarkable hierarchical structure is written as   
\be 
\label{6.10} 
\frac{m_{3}}{m_{2}} = \frac{m_{2}}{m_{1}}: \ m_{2}^{2} = m_{1}m_{3} \ \leftrightarrow \  
(e^{- \frac{1+2a}{6}\alpha})^{2} = e^{- \frac{3-2a}{6}\alpha}\cdot e^{- \frac{1-2a}{6}\alpha}.      
\ee 
We easily see that the condition is satisfied by a choice 
\be 
\label{6.11}
a = \frac{1}{4}. 
\ee 
By the way, in this case the mass ratio $m_{2}/m_{1} = e^{\frac{1}{6}\alpha} = e^{\frac{1}{2}\frac{\pi R_{5}}{R_{6}}}$. Interestingly. this prediction is the same with the corresponding prediction in the 2+1 model. It may be worth while noting that the observed slope of the $\log m_{f}$ plots for up-type, down-type quarks and charged lepton are all similar. Thus, there may be a possibility that this feature of our scenario may be relevant for the explanation of the observed fact. If we take, e.g., $m_{c} = 1.28$ GeV and $m_{u} = 2.8$ MeV and identify the ratio $\frac{m_{c}}{m_{u}}$ with $e^{\frac{1}{6}\alpha}$ to get a rough idea, we get $e^{- \alpha} = 1.1 \times 10^{-16}$ ($\alpha = 36.8 \ \to \ 
\frac{R_{5}}{R_{6}} = 3.9$). Thus the approximation made in (\ref{4.16}) seems to be reasonable, since $e^{- \alpha}$ is small enough, although $\frac{R_{5}}{R_{6}}$ itself is not so large, as the matter of fact. 

If we take into account the numerical factors appearing in (\ref{5.5}) seriously (replacing $j$ by $j+a$) to make (\ref{6.8}) more accurate, the double ratio is predicted to be (for $M = 3, \ a = \frac{1}{4}$)
\be 
\label{6.12} 
\frac{\left( \frac{m_{3}}{m_{2}} \right)}{\left( \frac{m_{2}}{m_{1}} \right)} 
= \frac{ \left( \frac{\frac{M^{2}}{4}-(1+a)^{2}}{\frac{M^{2}}{4} - (M-(2+a))^{2}} \right)} { \left( \frac{\frac{M^{2}}{4} - (M-(2+a))^{2}}{\frac{M^{2}}{4}-a^{2}} \right) } 
= \frac{ \left( \frac{11}{27} \right)}{ \left( \frac{27}{35} \right) } = 0.53.  
\ee 
On the other hand, if we take up-type quark masses ($m_{u} = 2.2$ MeV, $m_{c} = 1.28$ GeV, $m_{t} = 173$ GeV), for instance, the observed value of the double ratio is known to be 
\be 
\label{6.12'}
\frac{\left( \frac{m_{t}}{m_{c}} \right)}{\left( \frac{m_{c}}{m_{u}} \right)} = 0.23.  
\ee 
We have got a difference roughly of factor 2 again.

\subsection{Toward a realistic GHU model} 

So far we have been working on a 6D U(1)$\times$U(1) gauge theory. There left- and right-handed fermions have different quantum numbers, while the Yukawa coupling due to the Higgs field should be a sort of gauge interaction connection these two fermion fields in the GHU scenario. So we inevitably need non-Abelian gauge theory in order to realize the hierarchical fermion masses in this scenario. The simplest possibility for the non-Abelian gauge symmetry is SU(2). (Note that the minimal 5D unified electro-weak GHU model is SU(3) model \cite{Kubo:2001zc}, \cite{Scrucca:2003ra}.) In addition, the Abelian symmetry U(1)$_{1} \times$ U(1)$_{2}$ should be incorporated to the model.  
Thus, as a first step toward a realistic GHU model, here we consider SU(2)$\times$U(1) model, whose gauge group is of rank 2. A pair of fermions $\psi_{1}, \ \psi_{2}$ form a SU(2) doublet. Then the Higgs field connecting these two is associated with the off-diagonal generator of SU(2). The generators of U(1)$_{1}$ and U(1)$_{2}$, may be identified with two linear combinations, 
\be 
\label{7.1}
\frac{I + \tau_{3}}{2} 
= 
\left(
    \begin{array}{cc}
      1 & 0 \\
      0 & 0 
    \end{array}
  \right),  \ \ 
\frac{I - \tau_{3}}{2} 
= 
\left(
    \begin{array}{cc}
      0 & 0 \\
      0 & 1 
    \end{array}
  \right),
\ee 
where $\frac{\tau_{3}}{2}, \ \frac{I}{2} \ (I: \ {\rm unit \ matrix})$ are the third generator of SU(2) and the generator of U(1), respectively.  

Though we have discussed the effects of the magnetic monopole originally proposed by Dirac, it may also be an interesting possibility that the monopole proposed by 't Hooft-Polyakov \cite{'t Hooft} plays some role. In fact, their model is based on SO(3) gauge theory (isomorphic to SU(2)) and the introduced scalar field belongs to the adjoint representation of SO(3), with exactly the same feature also shared by the GHU scenario.  

Just as the orbifolding has an effect to reduce the original gauge symmetry, the presence of the magnetic monopole may cause a breakdown of gauge symmetry, SU(2) $\to$ U(1), since the pair of magnetic monopoles causes a non-trivial background field of $A_{6}$ associated with the 3rd generator $\frac{\tau_{3}}{2}$, which behaves as a sort of the vacuum expectation value of the adjoint scalar field from 4D point of view. 

The detailed discussions on these still unsettled issues of this model hopefully will be given in a separate paper.

\section{Summary discussion} 

The observed mass spectra of charged fermions show a remarkable hierarchical structure. Namely, if we plot the log of fermion masses as the function of ``generation number" for specific type of charged fermion, such as up-type quark, they seem to align along a straight line, roughly speaking. 
This seems to suggest us a great hint concerning the origin of fermion masses: the Yukawa couplings are originally all the same and universal for three generations and then get exponential suppression in some regular manner with ``quantized'' exponent, i.e. integer multiple of some unit quantity. 

In this paper we considered the possibility that such characteristic feature of the observed (charged) fermion mass spectra may be an inevitable consequence of the scenario of gauge-Higgs unification (GHU). The GHU is an attractive candidate of physics beyond the standard model, where Higgs field is identified with the extra space component of higher dimensional gauge field.  
In this scenario Yukawa couplings are all the same, i.e. gauge coupling constant, at least to start with, which provides a natural explanation why the Yukawa couplings are originally universal. In addition, the exponential suppression factor may be attributed to the corresponding suppression factor due to 
the ``$Z_{2}$-odd bulk mass", allowed to exist in 5-dimensional (5D) GHU model. 

The only remaining task is to find a natural mechanism to account for the quantized bulk mass, which is otherwise arbitrary parameter. In this paper we demonstrated that the bulk mass is naturally replaced by the background field of extra space component of gauge field, which originates from magnetic monopole placed inside of the torus as the extra dimension in a 6D GHU model. The size of 6th dimension was assumed to be much smaller than that of 5th dimension in order to acquire 5D viewpoint. Importantly, the quantized bulk mass is then realized by the well-known quantization condition of the magnetic charge imposed by Dirac. 

As the bonus, because of the presence of magnetic monopole, the theory turned out to become chiral, even if we do not adopt ``orbifolding". It also turned out that we have multiple Kaluza-Klein (KK) zero modes starting from just a single 6D Weyl fermion, which is useful to account for the generation structure.

To be concrete, we solved the differential equation for the mode function of KK zero mode under the influence of the magnetic monopole, by taking the quantization condition of the magnetic charge into account. Even though the equation was for a single 6D Weyl fermion, we obtained $M$ independent 4D Weyl fermions as the KK zero modes, where $M$ is an integer appearing in the quantization condition. Namely we could get $M$ generations starting from a single 6D field. 

Though the mode function of each chrality has exponential suppression factor and therefore shows localization at some point of the extra space, to get the desirable exponential suppression factor in the Yukawa coupling, localization in different points of the extra space depending on the chirality is needed and we had to introduce a pair of magnetic monopoles placed at opposite sides of the extra space. The magnetic monopoles are associated with independent U(1) symmetries, and the minimal framework to realize the desirable exponential suppression was argued to be U(1)$\times$U(1) gauge theory. 

After the overlap integral of the mode functions of left- and right-handed fermion we got the exponential suppression factor for the Yukawa coupling constant, as we expected. Then we investigated three generation model of our real interest, and demonstrated that there are two possibilities to realize the remarkable hierarchical structure of fermion mass spectrum: $2 + 1$ model and the model with twisted boundary condition. These (toy) models not only explain why the log of fermion mass is a linear function of generation number, but also predict slight deviation from this relation, which has the same tendency with what the observed fermion masses show. The agreement between theoretical prediction and the observed value concerning the double ratio of three mass eigenvalues $(m_{3}/m_{2})/(m_{2}/m_{1})$ is reasonable. We also argued that these two models predict the same slope in the plot of the $\log m_{f}$ as the function of the generation number. It may be interesting, on the other hand, to note that the observed slope for up-type, down-type quarks and charged lepton are all similar. 

There still remain many issues to be settled in this approach to explain the remarkable hierarchical fermion mass spectrum in the framework of GHU. For instance, in the model with twisted boundary condition, we need a reason why the parameter $a$ takes such specific value $1/4$. Since this parameter describes the twisted boundary condition along the cycle of 6th dimension, it may be replaces by the effect of a sort of Wilson-loop phase. Then, it may be fixed by the minimization of the radiatively induced effective potential with respect to the phase, as we usually perform in the GHU models. Also, as was discussed in the last subsection, there still remain several unsettled issues in the attempt to construct a realistic model, such as how non-Abelian gauge symmetry is broken by the presence of the magnetic monopole, the relevance of the 't Hooft-Polyakov monopole, etc. 

Though the presence of magnetic monopole was just assumed in this paper, its origin may be attributed to some solitonic object in the string theory, such as brane. Finally a comment on the neutrino mass is in order. In this paper we discussed the implication of the GHU scenario for the mass spectrum of charged fermion. But, the scenario may have an interesting implication also for the neutrino mass. We have seen that if we introduce a pair of fermions with $M = 3$ and adopt periodic boundary condition along the cycle of 6th dimension, there appear three generations of fermion, but with degenerate mass eigenvalues, roughly  behaving as $e^{- \frac{\pi R_{5}}{R_{6}}\frac{3}{2}}, \ e^{- \frac{\pi R_{5}}{R_{6}}\frac{1}{2}}, \ e^{- \frac{\pi R_{5}}{R_{6}}\frac{1}{2}}$ ($R_{5, 6}$: the size of 5th and 6th dimension, respectively). Though this clearly contradicts with the observation for charged fermions, when the mechanism is applied to the neutrino mass, it seems to imply the ``inverted hierarchy" scenario of neutrino masses, though the degeneracy between higher two mass eigenvalues has to be resolved by some minor effect (such as radiative correction ?).   

Hopefully, these still unsettled issues will be discussed in a separate paper.      
 
\subsection*{Acknowledgments}

The author would like to thank T. Kobayashi for informative and very valuable discussions in the early stage of this work. This work was supported in part by Japan Society for the Promotion of Science, Grants-in-Aid for Scientific Research, No.~16H00872, No.~15K05062. 


\providecommand{\href}[2]{#2}\begingroup\raggedright\endgroup

\end{document}